# Photoelectrical detection of electron spin resonance of nitrogen-vacancy centres in diamond

E. Bourgeois[1,2], A. Jarmola[3], M. Gulka[1,4], J. Hruby[1,4], D. Budker[3,5] and M. Nesladek[1,2]*

[1]*IMOMEC division, IMEC, Kapeldreef 75, B 3001 Leuven, Belgium*
[2] *Hasselt University, Martelarenlaan 42, B-3500 Hasselt, Belgium*
[3]*Department of Physics, University of California, Berkeley, California 94720-7300, USA*
[4]*Czech Technical University in Prague, Sitna sq. 3105, 272 01, Kladno, Czech Republic*
[5]*Helmholtz Institute, Johannes Gutenberg University, Mainz, Germany*

*Correspondence to: milos.nesladek@uhasselt.be

**Abstract**:

The protocols for the control and readout of Nitrogen Vacancy (NV) centres electron spins in diamond offer an advanced platform for quantum computation, metrology and sensing. These protocols are based on the optical readout of photons emitted from NV centres, which process is limited by the yield of photons collection. Here we report on a novel principle for the detection of NV centres magnetic resonance in diamond by directly monitoring spin-preserving electron transitions through measurement of NV centre related photocurrent. The demonstrated direct detection technique offers a sensitive way for the readout of diamond NV sensors and diamond quantum devices on diamond chips. The Photocurrent Detection of Magnetic Resonance (PDMR) scheme is based on the detection of charge carriers promoted to the conduction band of diamond by the two-photon ionization of $NV^-$ centres. Optical detection of magnetic resonance (ODMR) and PDMR are compared, by performing both measurements simultaneously. The minima detected in the measured photocurrent at resonant microwave frequencies are attributed to the spin-dependent occupation probability of the $NV^-$ ground state, originating from spin-selective non-radiative transitions.

**One Sentence Summary:**

We report novel photoelectrical readout of NV centres magnetic resonance in diamond and describe the mechanism allowing this novel detection scheme.

**Main Text:**

Addressing and reading the electron spin of nitrogen-vacancy (NV) centres in diamond hold great promise for quantum computing and secure communication (*1*, *2*), as well as for nanoscale magnetic and electric sensing (*3*, *4*, *5*) and for non-perturbing sensing and imaging of quantum objects (*6*, *7*). At present, the readout of NV centres electron spin state is only performed optically, by detecting photons using confocal microscopy. No simple spin read-out techniques allowing for integration with electronic chips are available. One of the limiting factors of the optical detection scheme is the low collection efficiency of photons emitted by NV centres in bulk diamond. Due to the limitations of objective optics and to the diamond index of refraction, detection efficiency higher than a few percent can only be obtained by complex micro-fabrication (*8*). Additionally, NV centres-based quantum computing involves positioning adjacent NV centres at a distance of approximately 10 nm (*9*). The individual optical readout of such close NV centres is difficult, since it requires a resolution below the diffraction limit. In this report we describe a novel non-optical technique based on the direct photoelectrical read-out of NV centres electron spin resonance. Contrary to a recently proposed scheme, in which the read-out of NV centres is performed via the monitoring of non-radiative energy transfers to graphene (*10*), our method relies on the direct photo-excitation of charge carriers from NV centres to the diamond conduction band and on the collection of charge carriers by electrodes fabricated on the diamond chip. This scheme is denoted as photocurrent detection of magnetic resonance (PDMR). The presented technique avoids the complexity of confocal imaging and allows the detection of NV spin resonance in light scattering media, which will be an advantage for various sensing applications. The PDMR technique, demonstrated here on spin ensembles, can be used for the readout of individual NV centres spins. It may in addition significantly outperform the efficiency of optical detection since every photon has the ability to generate more than one electron-hole pair, due to the photocurrent gain mechanism (*11*).

The detailed mechanism for the photoelectrical detection of NV centre electron spin resonance is reported below. To demonstrate this detection principle, optical detection of magnetic resonance (ODMR) and PDMR are performed simultaneously on a type-Ib single-crystal diamond plate of [100] crystallographic orientation. A crystal with an initial concentration of approximately 200 ppm of substitutional nitrogen ($N_s^0$) is electron-irradiated and thermally annealed, leading to a concentration of $NV^-$ centres around 20 ppm. Coplanar interdigitated electrodes are prepared on the oxidized diamond surface. A green linearly polarized laser beam (532 nm) pulsed by an acousto-optical modulator is focused onto the diamond surface, in between the coplanar electrodes. Photoluminescence light emitted under the effect of green excitation is collected, filtered and focused onto a pyroelectric detector connected to a lock-in amplifier. For photocurrent measurements, the photo-carriers generated upon green illumination are driven towards electrodes by an electric field, created by applying a DC voltage between electrodes [Fig. 1(A)]. The photocurrent is amplified with a low-noise current preamplifier and measured using a lock-in amplifier. Lock-in amplification leads to a high signal-to-noise ratio and allows us to detect a photocurrent in the range of 10 to 100 pA with a four digit precision. A microwave field of fixed power is produced using a metal wire pressed across the diamond surface, and insulated from coplanar electrodes by a coating (see Supplementary materials for details on sample preparation and experimental set-up).

The intensity of photoluminescence and photocurrent are measured simultaneously while scanning the microwave frequency, in the absence and in the presence of an external magnetic field [Fig. 1(B)]. Minima in the photoluminescence intensity are observed at microwave frequencies inducing resonant transitions between the |0> and the |±1> spin sublevels of the $NV^-$ spin triplet ground state ($^3A_2$), as classically reported in ODMR measurements (*12*). Minima in photocurrent are clearly detected at identical frequencies, demonstrating that photocurrent measurements can be used to detect the spin resonances of $NV^-$ centres.

The two resonances observed in ODMR and PDMR spectra in the absence of an external magnetic field indicate the existence of a zero-field splitting between linear combinations of the |+1> and |-1> spin sublevels of $NV^-$, induced by local strain in the material (*13*). The origin of the small difference (< 1.5 MHz) in the position of the minima observed in ODMR and PDMR spectra is discussed in Supplementary materials. In the presence of an external static magnetic field applied using a permanent magnet, a further splitting of resonant frequencies is observed, both in ODMR and PDMR spectra, reflecting the shift of |-1> and |+1> spin sublevels of $NV^-$ ground state due to the Zeeman effect. As expected (14), two magnetic resonances are observed when the magnetic field is applied along the [100] direction of the diamond crystal and four when it is applied along the [111] direction [Fig. 1(B)].

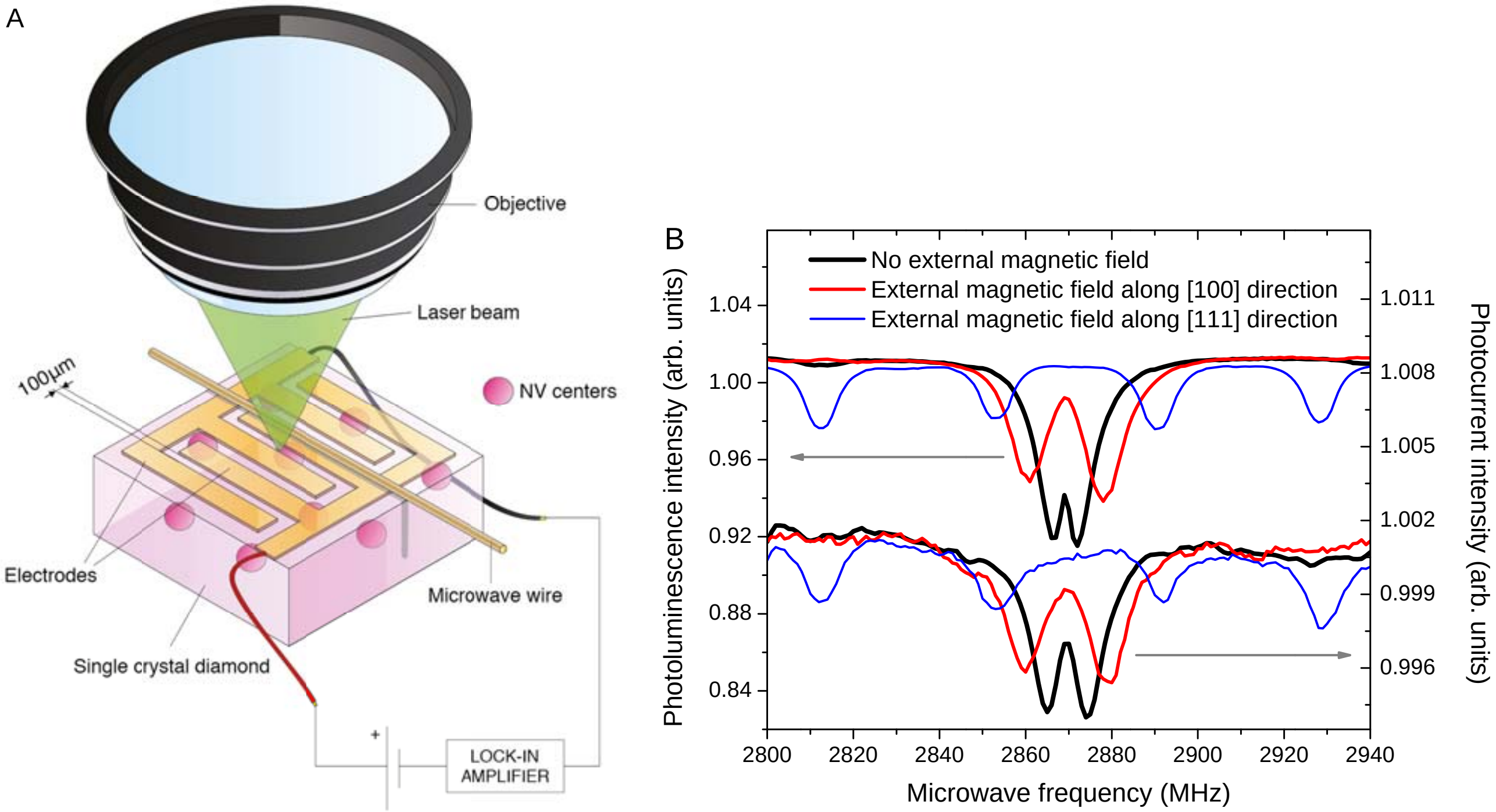

**Fig. 1:** Schematic representation of the set-up used for PDMR (**A**). Comparison of ODMR (top part of the figure, left axis) and PDMR (bottom part of the figure, right axis) spectra recorded simultaneously (green light power: 100 mW) (**B**). Spectra are normalized for clarity.

To determine the photo-ionization mechanism inducing the photocurrent minima at NV$^-$ spin resonance frequencies, the photocurrent intensity has been measured as a function of the green light power (Fig. 2). A good fit to the experimental data is obtained using the sum of a linear and a quadratic function, indicating that the measured photocurrent results from the superposition of a one-photon and a two-photon ionization process (*15*). In previous experiments dealing with NV centre physics (*15*) and in theoretical studies (*16, 17*) it has been demonstrated that a photon energy higher than 2.6 eV was necessary to induce the photo-ionization of NV$^-$ via a one-photon process, i.e. to directly promote an electron from the ground state of NV$^-$ to the conduction band. Based on this argument and on the quadratic power dependence of the photocurrent, we conclude that a two-photon absorption process is responsible for the NV-related part of the photocurrent induced by green light (2.33 eV). The mechanism for the two-photon ionization of NV$^-$ centres in diamond has been experimentally established (*15, 18, 19*) and modelled (*19*). In this process, the absorption of a first photon promotes an electron from the $^3A_2$ triplet ground state of NV$^-$ to its $^3E$ triplet excited state [Fig. 3(A), transition (1)] and a second photon excites this electron to the conduction band of diamond [Fig. 3(B), (4)], which results in the conversion of the NV centre to its neutral state NV$^0$ and in the promotion of an electron into the conduction band [Fig. 3(C)]. To ensure the charge neutrality and the photocurrent stability in the PDMR detection circuit, the NV$^0$ centres formed by ionisation of NV$^-$ have to be subsequently converted back to the NV$^-$ state, either by capturing an electron from a donor defect (in particular from N$_s^0$, present in high concentration in the 1b sample under study) (*20, 21)* or by two-photon conversion from NV$^0$ to NV$^-$. In the latter process, the absorption of a first photon excites the NV$^0$ centre [Fig. 3(C), (5)], while a second photon promotes an electron from the valence band to the vacated orbital of NV$^0$ [Fig. 3(C), (6)], leaving a hole in the valence band (*15, 19*).

The linear fraction of the measured photocurrent (Fig. 2) is most probably associated with the one-photon ionization of $N_S^0$. $N_S^0$ is indeed the dominant point defect in our sample and the threshold energy for the photoionization of this defect is well below the 2.33 eV excitation energy used in our experiment, as reported in (*22*, *23*).

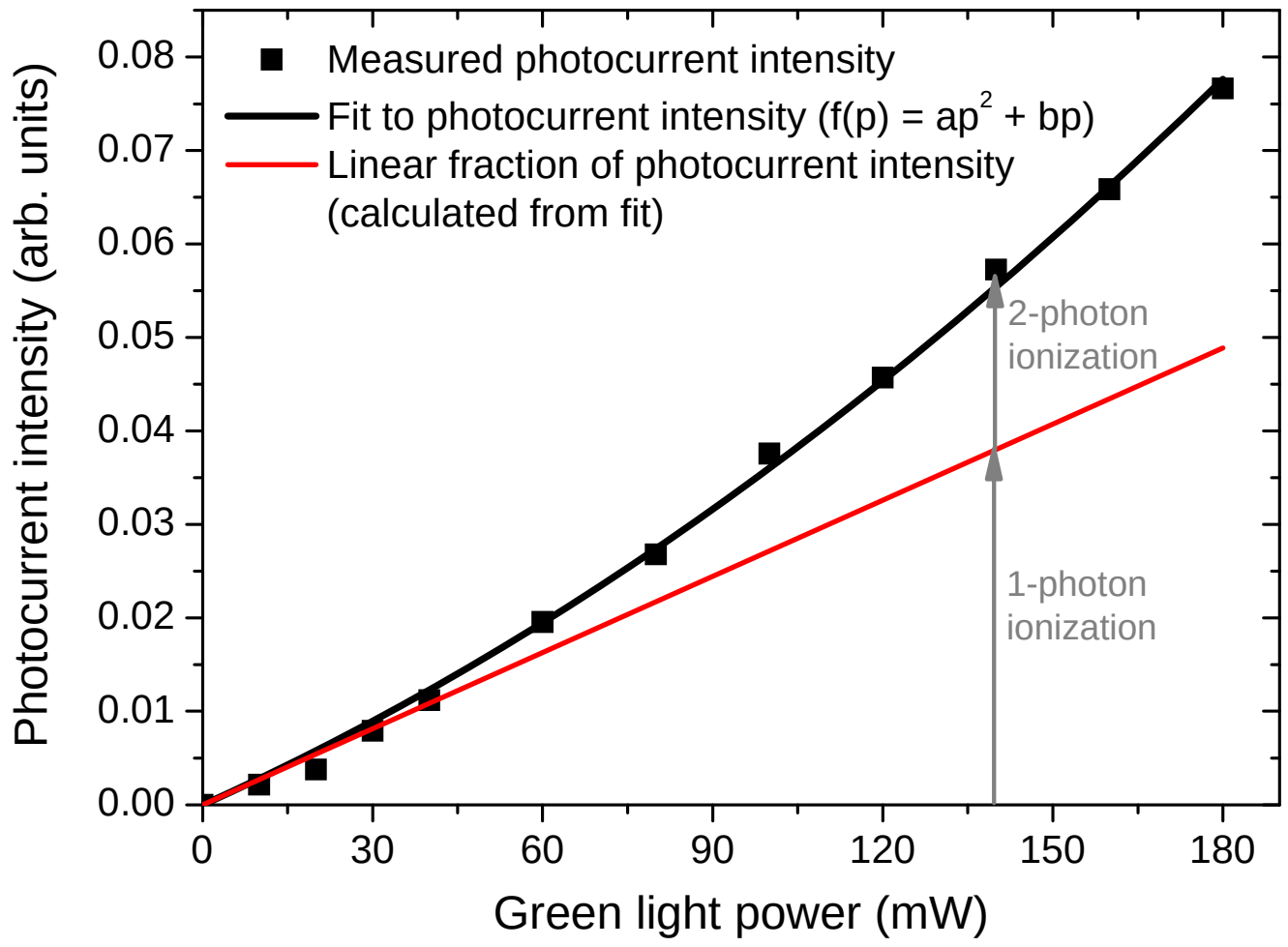

**Fig. 2:** Influence of the 532 nm light power on the photocurrent intensity.

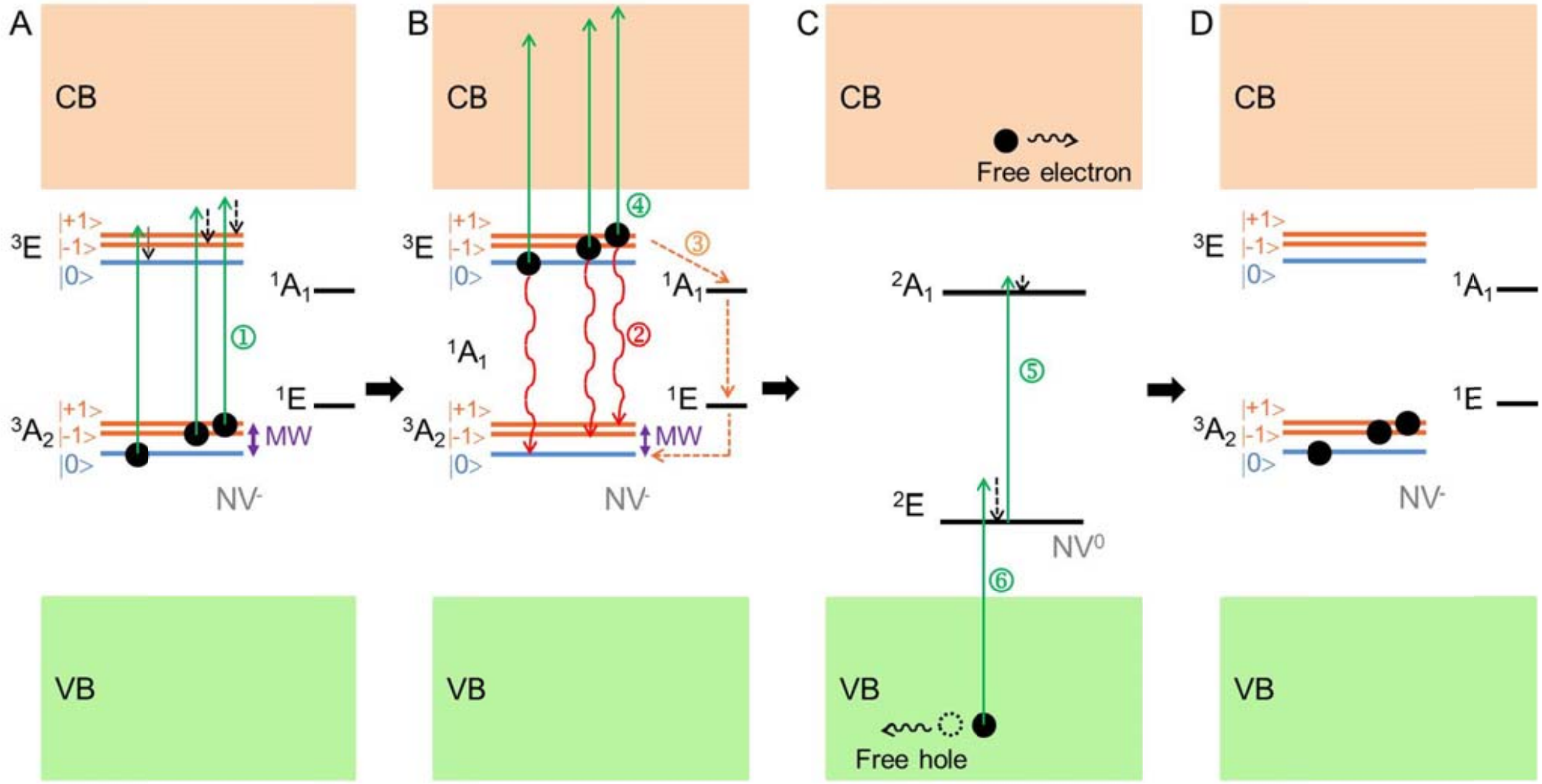

**Fig. 3:** Energy level scheme of the $NV^-$ centre and description of the mechanism proposed to explain PDMR measurements. For the purpose of illustration, the distance between the top of the diamond valence band (VB) and the ground state of $NV^-$ is not represented at the correct scale. The absorption of a first photon promotes an electron from the $^3A_2$ ground state to the $^3E$ excited state of $NV^-$ (**A**, transition (1)). From $^3E$, electrons can radiatively decay to $^3A_2$ (2) or be excited to the diamond conduction band (CB) by the absorption of a second photon (4). The spin-selective non-radiative decay of electrons to the metastable singlet state $^1A_1$ (3) enables PDMR (**B**). Two-photon ionization of $NV^-$ results in the formation of a $NV^0$ centre and a free electron in the CB (**C**). By a two-photon process (5) (6) the NV centre can finally be converted back to its negatively charged state (**D**).

Figure 1(B) confirms that our PDMR detection scheme allows for the detection of $NV^-$ spin resonance, by monitoring the spin-preserving transitions. The minima observed in photocurrent correspond to resonant MW frequencies inducing transitions from the |0> to the |±1> electron spin sublevels of $NV^-$ ground state ($^3A_2$). From this observation, it can be inferred that electrons in the |±1> sublevels have a lower probability to be promoted to the conduction band than electrons initially situated in the |0> spin sublevel. In ODMR, the contrast in photoluminescence intensity between $NV^-$ |0> and |±1> spin sublevels originates from the existence of the singlet states ($^1A_1$ and $^1E$) in the electronic structure of $NV^-$ (*24*). Under the effect of green illumination, electrons are initially pumped into the $^3A_2$ |0> spin state, from which they are coherently driven to the $^3A_2$ |±1> spin sublevels by the resonant microwave field (*25*). Since optical transitions between $NV^-$ ground and excited state are spin conserving, the absorption of a photon promotes electrons initially in the |0> (respectively |±1>) spin sublevel of $^3A_2$ to the |0> (respectively |±1>) spin sublevel of $^3E$ [Fig. 3(A), (1)]. Photoluminescence originates from the radiative decay of electrons to the ground state [Fig. 3(B), (2)]. However, due to spin-selective intersystem crossing (*26*), electrons in the $^3E$ |±1> sublevels have a much higher probability than electrons in the $^3E$ |0> state to decay non-radiatively to the $^1A_1$ singlet state [lifetime ≤ 1 ns (*24*)], from which they fall to the $^1E$ metastable singlet state [Fig. 3(B)]. This non-radiative decay path [Fig. 3(B), (3)] induces a difference between the photoluminescence brightness of the different $NV^-$ spin sublevels and enables ODMR. In the same way, the photocurrent detection of magnetic resonances is enabled by the existence of this non-radiative decay path. Since the $^3E$ |±1> spin sublevels present a shorter lifetime (7.3 ns) than the $^3E$ |0> sublevel (13.7 ns) (*27*), an electron excited to the $^3E$ |±1> spin sublevels has a lower probability to be promoted to the conduction band by the absorption of a second photon [Fig. 3(B), (4)] than an electron occupying the $^3E$ |0> sublevels. In addition, for the time during which the $^3E$ |±1> electron undergoes the cycle of optical transitions via inter-state crossing to the $^3A_2$ |0> sublevel, it does not contribute to the photocurrent. The $^1E$ metastable state has a lifetime of 220 ns at room temperature (*24*). For that period, the $^1E$ state stores the electron, leading to a temporary decrease in the occupation of $NV^-$ ground state and reducing the rate of two-photon ionization (proportional to the occupation of $^3A_2$).

Figure 4 depicts the influence of the green light power on the ODMR and PDMR contrasts. The maximal observed contrast is ~1 % for PDMR and 10.7 % for ODMR. This is to a large degree due to the fact that for the measured sample only the two-photon fraction of the photocurrent (associated to the ionization of $NV^-$) gives rise to detectable electron spin resonances, while its linear fraction is not affected by the microwaves. The monotonous increase in the PDMR contrast with the incident light power (Fig. 4) can thus be explained by the increase in the two-photon quadratic fraction of the photocurrent with respect to the linear fraction. However, while the quadratic fraction of the photocurrent increases by a factor 4 between 30 mW and 180 mW (Fig. 2), the PDMR contrast only increases by a factor 2. This can be partly explained by the fact that in the same range of laser power the ODMR contrast decreases (Fig.4), which has been attributed to saturation (*14*).

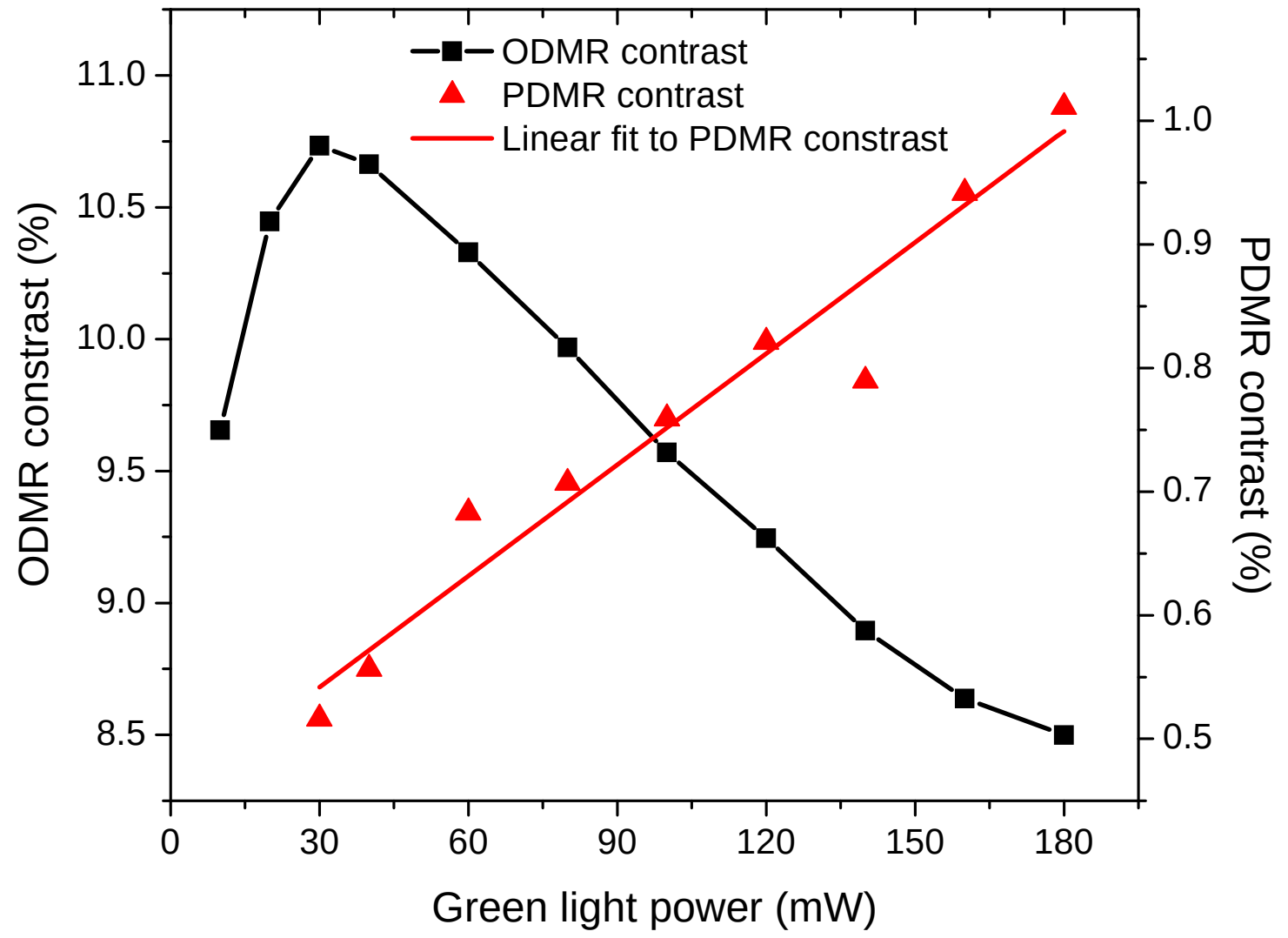

**Fig. 4:** Influence of the 532 nm light power on the ODMR and PDMR contrasts [defined as the depth of the drop in the photoluminescence or photocurrent signal at the second zero-magnetic field resonance ($F_2$ ~ 2873 MHz)].

An increase in the PDMR contrast could be obtained by further increasing the power of the green excitation, or by using photons of lower energy to induce the ionization of $NV^-$ centres without ionizing $N_s^0$ or other defects in diamond.

The presented experiment demonstrates a new principle for the readout of the magnetic resonance of NV centres electron spins in diamond. By reducing the contact area, this technique has the potential to address single NV centres. This paradigm may lead to a sensitive way for the construction of diamond nanoscale sensors and quantum devices and their direct readout, allowing directly performing quantum operation on chip.

**Acknowledgments:** This work was supported by the AFOSR/DARPA QuASAR program, The Research Foundation - Flanders (FWO) (Grant agreement G.088812N), the European Union Seventh Framework Programme (FP7/2007-2013) under the project DIADEMS (grant agreement n°611143) and the Ministry of Science and Education of the Czech Republic (Contract CZ.1.07/2.3.00/20.0306). The authors are grateful to F. Jelezko and P. Kehayias for helpful discussions, to K. Aulenbacher and the MAMI staff for electron irradiation, and to Y. Balasubramaniam and P. Robaeys for electrodes fabrication.